\documentclass[epj,twocolumn,applemac]{webofc}
\usepackage[varg]{txfonts}   
\woctitle{Hadron Collider Physics symposium 2012}
\begin{document}
\title{ALICE Physics Summary}
%
%

\author{Roberto Preghenella\inst{1,2}\fnsep\thanks{\email{preghenella@bo.infn.it}} for the ALICE Collaboration}

\institute{Centro Studi e Ricerche e Museo Storico della Fisica “Enrico Fermi”, Rome, Italy 
  \and
  Sezione INFN, Bologna, Italy
}

\abstract{%
  
  The ALICE experiment at the LHC has collected data in proton-proton (pp) collisions at $\sqrt{s}$ = 0.9, 2.76, 7 and 8 TeV, in lead-lead (Pb--Pb) collisions at $\sqrt{s_{\rm NN}}$ = 2.76 TeV and in proton-lead (p-Pb) collisions at $\sqrt{s_{\rm NN}}$ = 5.02 TeV. A summary overview of recent experimental physics results obtained by ALICE is presented in this paper with a selection of few representative measurements.
  
}
\maketitle
\section{Introduction}
\label{intro}

ALICE (A Large Ion Collider Experiment) is a general-purpose heavy-ion detector at the CERN LHC (Large Hadron Collider) whose design has been chosen in order to fulfill the requirements to track and identify particles from very low ($\sim$100~MeV/$c$) up to quite high ($\sim$100~GeV/$c$) transverse momenta in an environment with large charged-particle multiplicities as in the case of central lead-lead (Pb--Pb) collisions at extremely high energies.

\begin{figure}[b]
\includegraphics[width=\linewidth,clip]{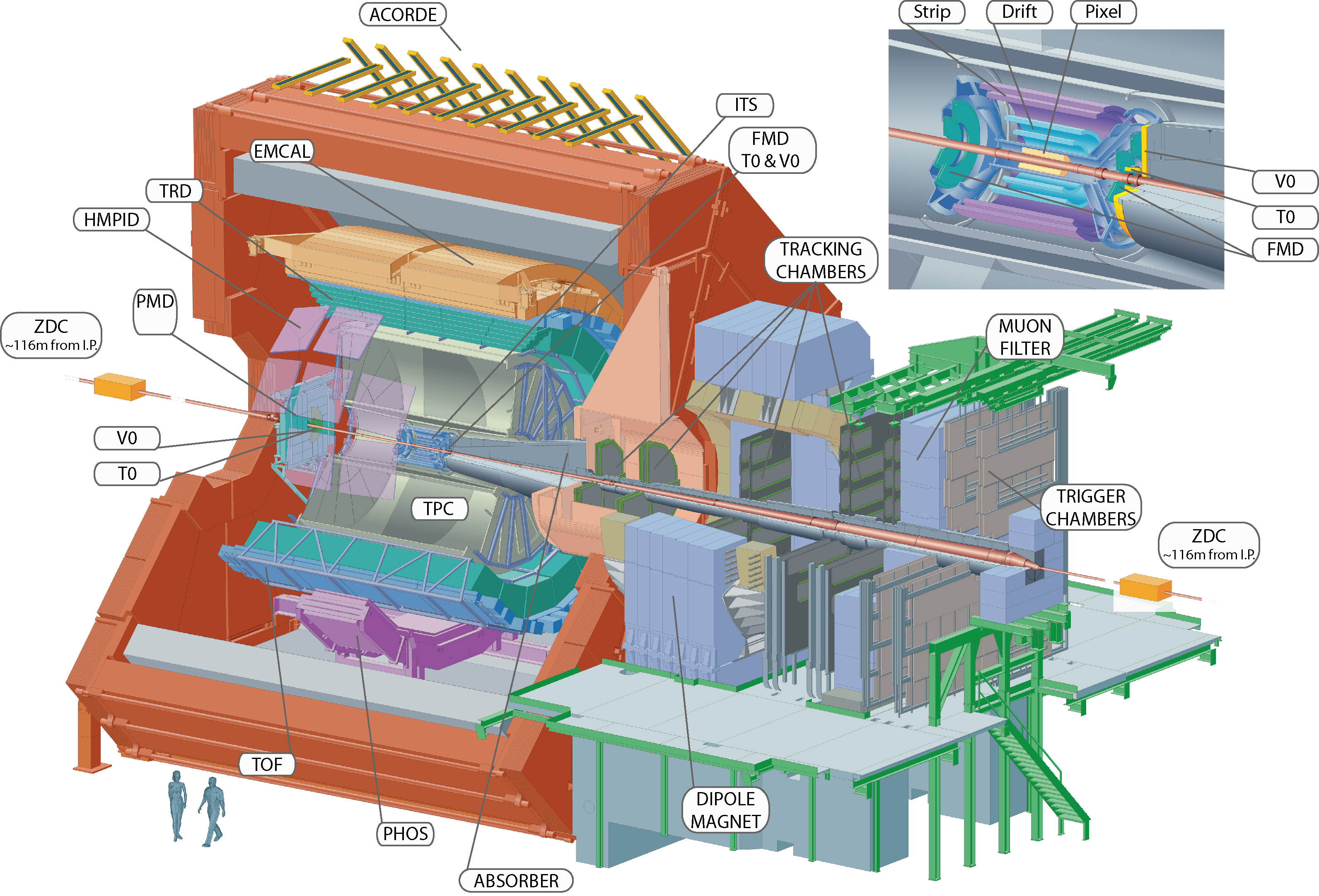};
\caption{Schematic layout of the ALICE detector with its main subsystems.}
\label{fig:alice}
\end{figure}

The ALICE experiment, shown in Figure~\ref{fig:alice}, consists of a central-barrel detector system and several forward detectors.
The central system covers the mid-rapidity region ($\left| \eta \right| \leq$ 0.9) over the full azimuthal angle and it is installed inside a large solenoidal magnet with a moderate magnetic of 0.5 T. It includes a six-layer high-resolution inner-tracking system (ITS), a large-volume time-projection chamber TPC) and electron and charged-hadron identification detectors which exploit transition-radiation (TRD) and time-of-flight (TOF) techniques, respectively. Small-area detectors for high-$p_{\rm T}$ particle-identification (HMPID), photon and neutral-meson measurements (PHOS) and jet reconstruction (EMCal) complement the central barrel.
The large-rapidity systems include a single-arm muon spectrometer covering the pseudorapidity range -4.0 $\leq \eta \leq$ -2.4 and several smaller detectors (VZERO, TZERO, FMD, ZDC, and PMD) for triggering, multiplicity measurements and centrality determination. A detailed description of the ALICE detector layout and of its subsystems can be found in~\cite{Aamodt:2008zz}.

Since November 2009 when the first LHC collisions occured, the ALICE detector collected proton-proton data at several centre-of-mass energies ($\sqrt{s}$ = 0.9, 2.76, 7 and 8 TeV). During the first two LHC heavy-ion runs, in the falls of 2010 and 2011, the ALICE experiment recorded Pb--Pb collisions at a centre-of-mass energy per nucleon pair of $\sqrt{s_{\rm NN}}$ = 2.76 TeV and could profit of an integrated luminosity of about 10 $\mu \rm b^{-1}$ and 100 $\mu \rm b^{-1}$, respectively. The instant luminosity exceeded $10^{26}\rm cm^{-2} s^{-1}$ in the second run, higher than the design value. 
Proton-lead (p--Pb) collision data were also recorded with the ALICE detector. This occured during a short run performed in September 2012 in preparation for the main p--Pb run scheduled at the beginning of 2013. Each beam contained 13 bunches; 8 pairs of bunches were colliding in the ALICE interaction region, providing a luminosity of about $8 \times 10^{25}\rm cm^{-2}s^{-1}$. Beam 1 consisted
of protons at 4 TeV energy circulating in the negative z-direction in the ALICE laboratory system, while beam 2 consisted of fully stripped $^{208}$Pb ions at $82 \times 4$ TeV energy circulating in the positive z-direction. This configuration resulted in collisions at $\sqrt{s_{\rm NN}}$ = 5.02 TeV in the nucleon-nucleon centre-of-mass system, which moves with a rapidity of $\Delta y_{\rm NN}$ = 0.465 in the direction of the proton beam.

\section{First results in proton-lead collisions}
\label{sec-1}
Particle production in proton-lead (p--Pb) collisions allows to study and understand QCD at low parton fractional momentum $x$ and high gluon density. Moreover it is expected to be sensitive to nuclear effects in the initial state. For this reason p--Pb measurements provide an essential reference tool to discriminate between initial and final state effects and they are crucial for the studies and the understanding of deconfined matter created in nucleus-nucleus collisions.

\subsection{Charged-particle pseudorapidity density}

\begin{figure}[t]
\includegraphics[width=\linewidth,clip]{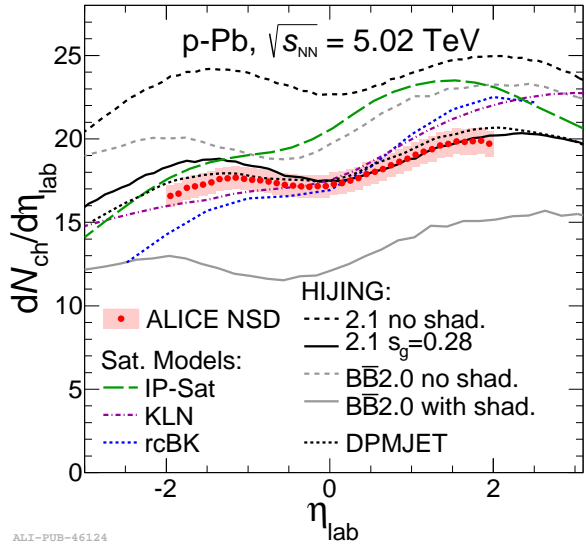}
\caption{Pseudorapidity density of charged particles measured in NSD p--Pb collisions at $\sqrt{s_{\rm NN}}$ = 5.02 TeV compared to theoretical predictions.}
\label{fig:dndeta}
\end{figure}

The measurement of primary charged-particle pseudorapidity density performed in non-single diffractive (NSD) p--Pb collisions at $\sqrt{s_{\rm NN}}$ = 5.02 TeV is reported in \cite{ALICE:2012xs}. The resulting pseudorapidity density is presented in Figure~\ref{fig:dndeta} as measured in the laboratory system for $\left| \eta_{\rm lab} \right| < 2$. A forward-backward asymmetry between the proton and the lead hemispheres is clearly visible. The data is also compared to several model predictions of particle production which have been shifted in the laboratory system when needed. The comparison shows that the pseudorapidity dependence is best described by the models DPMJET and HIJING 2.1 (where the gluon shadowing parameter $s_{g}$ was tuned on experimental $\sqrt{s_{\rm NN}}$ = 200 GeV d--Au data at RHIC), whereas the saturation models (KNL, rcBK, IP-Sat) exibit a steeper $\eta_{\rm lab}$ dependence. The pseudorapidity density in the centre-of-mass system at mid-rapidity $\left| \eta_{\rm cms} \right| < 0.5$ is ${\rm d}N_{\rm ch}/{\rm d}\eta_{\rm cms} = 16.81 \pm 0.71$, corresponding to $2.14 \pm 0.17$ charged particles per unit pseudorapidity per participant when scaled by the number of participating nucleons, determined using the Glauber model~\cite{Alver:2008aq}.

\subsection{Transverse momentum distribution}

\begin{figure}[t]
\includegraphics[width=\linewidth,clip]{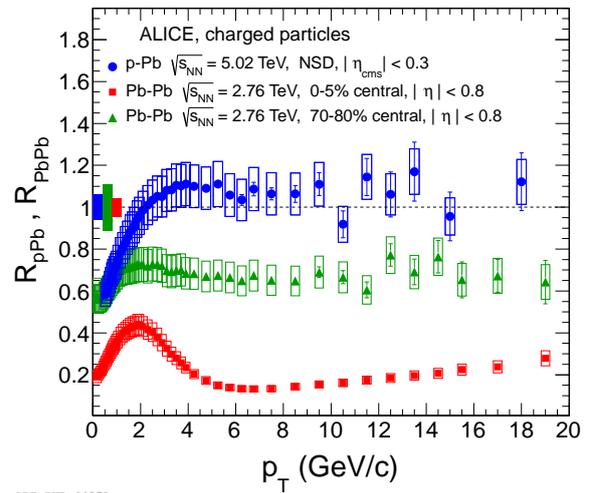}
\caption{The nuclear modification factor of charged particles as a function of transverse momentum in NSD p--Pb collisions at $\sqrt{s_{\rm NN}}$ = 5.02 TeV compared to measurements in central (0-5\%) and peripheral (70-80\%) Pb--Pb collisions at $\sqrt{s_{\rm NN}}$ = 2.76 TeV.}
\label{fig:rppb}
\end{figure}

The measurement of the transverse momentum $p_{\rm T}$ distributions of charged particle in p--Pb collisions were also reported~\cite{ALICE:2012mj}. It was previously shown that the production of charged hadrons in central Pb--Pb collisions at the LHC is strongly suppressed~\cite{Aamodt:2010jd,:2012eq}. The suppression remains substantial up to 100 GeV/$c$ and is also seen in reconstructed jets~\cite{:2012is}. Proton-lead collisions provide a control experiment to establish whether the initial state of the colliding nuclei plays a role in the observed high-$p_{\rm T}$ hadron production in Pb--Pb collisions. In order to quantify nuclear effects, the $p_{\rm T}$-differential yield relative to the proton-proton reference, the so-called nuclear modification factor, is calculated. The nuclear modification factor is unity for hard processes which are expected to exhibit binary collision scaling. This has been recently confirmed in Pb--Pb collisions at the LHC by the measurements of direct photon, Z$^{0}$ and W$^{\pm}$ production, observables which are not affected by hot QCD matter. In Figure~\ref{fig:rppb} the measurement of the nuclear modification factor in p--Pb collisions R$_{\rm pPb}$ is compared to that in central (0-5\% centrality) and peripheral (70--80\%) Pb--Pb collisions R$_{\rm PbPb}$. R$_{\rm pPb}$ is observed to be consistent with unity for transverse momenta higher that about 2 GeV/$c$. This demonstrates that the strong suppression observed in central Pb--Pb collisions at the LHC is not due to an initial-state effect, but it is rather a final state effect related to the hot matter created in high-energy heavy-ion collisions. 

\section{Recent physics results}
\label{sec-1}
A limited selection of recent physics results obtained by the ALICE experiment at the LHC is reported in the following section. They include both measurements performed in proton-proton (pp) and in lead-lead (Pb--Pb) collisions and are meant to give a feeling of the physics capabilities of the experiment, though this cannot be done in a complete way in this report.

\subsection{Exclusive $\rm J/\psi$ photoproduction}

\begin{figure}[t]
\includegraphics[width=\linewidth,clip]{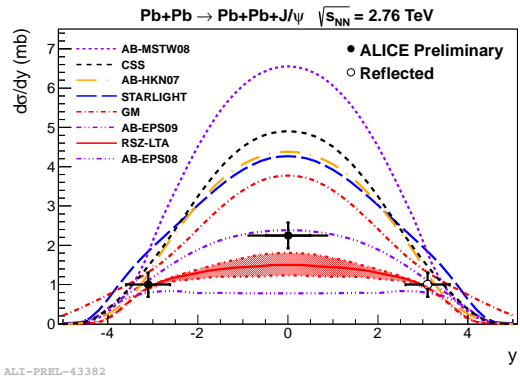}
\caption{Measured differential cross section of coherent $\rm J/\psi$ photoproduction compared with models. The point at positive rapidity is the reflected result from the measurement at negative rapidities.}
\label{fig:ultraper}
\end{figure}

Exclusive vector meson production in heavy-ion interactions is expected to probe the nuclear gluon distribution for which there is considerable uncertainty in the low-$x$ region and it has been studied so far in gold-gold (Au--Au) collisions at RHIC. The first LHC results on exclusive photoproduction of $\rm J/\psi$ vector mesons measured at forward rapidities in ultra-peripheral Pb--Pb collisions at $\sqrt{s_{\rm NN}}$ = 2.76 TeV has been reported here~\cite{Abelev:2012ba}. The forward and mid-rapidity measured differential cross section for coherent $\rm J/\psi$ production is compared with calculations from various models in Figure~\ref{fig:ultraper}. Best agreement is found with models which include nuclear gluon shadowing consistent with the EPS09 parametrizations of the nuclear gluon distribution functions (AB-EPS09).

\subsection{Direct photons}

\begin{figure}[t]
\includegraphics[width=\linewidth,clip]{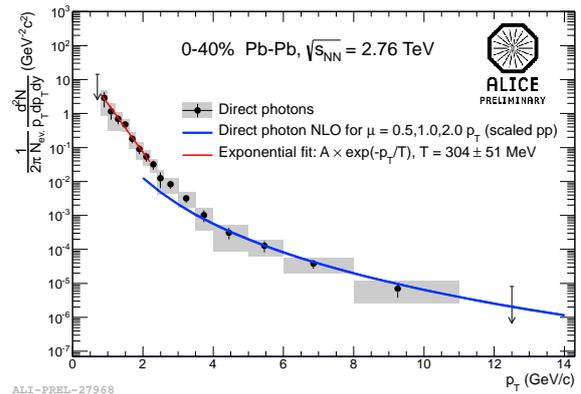}
\caption{Direct-photon invariant yield in 0-40\% central Pb--Pb collisions at $\sqrt{s_{\rm NN}}$ = 2.76 TeV with NLO pQCD prediction and exponential fit in low-pT region.}
\label{fig:directgamma}
\end{figure}

Direct-photon production has been measured by ALICE in proton-proton and in Pb--Pb collisions~\cite{Wilde:2012wc}. The measurements when compared to NLD pQCD predictions are found to be in good agreement for pp collisions and peripheral Pb--Pb collisions (40-80\% centrality). On the other hand, in the case of central (0-40\% centrality) Pb--Pb collisions the direct photon signal is well reproduced by NLO pQCD only for photon momenta above 4 GeV/$c$, as shown in Figure~\ref{fig:directgamma}. The low-$p_{\rm T}$ excess, of about 20\% at around 2 GeV/$c$, is attributed to thermal photons, that is photons produced in the QGP phase by the scattering of thermalized partons. The thermalized nature of the production medium should be reflected in the $p_{\rm T}$ distribution of thermal photons. The excess over NLO pQCD is fit in Figure~\ref{fig:directgamma} with an exponential in the $p_{\rm T}$ range 0.8--2.2 GeV/$c$. The inverse slope of this exponential is found to be T = (304 $\pm$ 51) MeV. In a similar analysis performed in central (0-20\%) Au--Au collisions at $\sqrt{s_{\rm NN}}$ = 200 GeV, the PHENIX experiment at RHIC measures an inverse slope parameter of T = (221 $\pm$ 19 $\pm$ 19) MeV. The LHC value is about 40\% higher than the one measured at RHIC.

\subsection{Light-flavour hadron production}

\begin{figure}[t]
\includegraphics[width=\linewidth,clip]{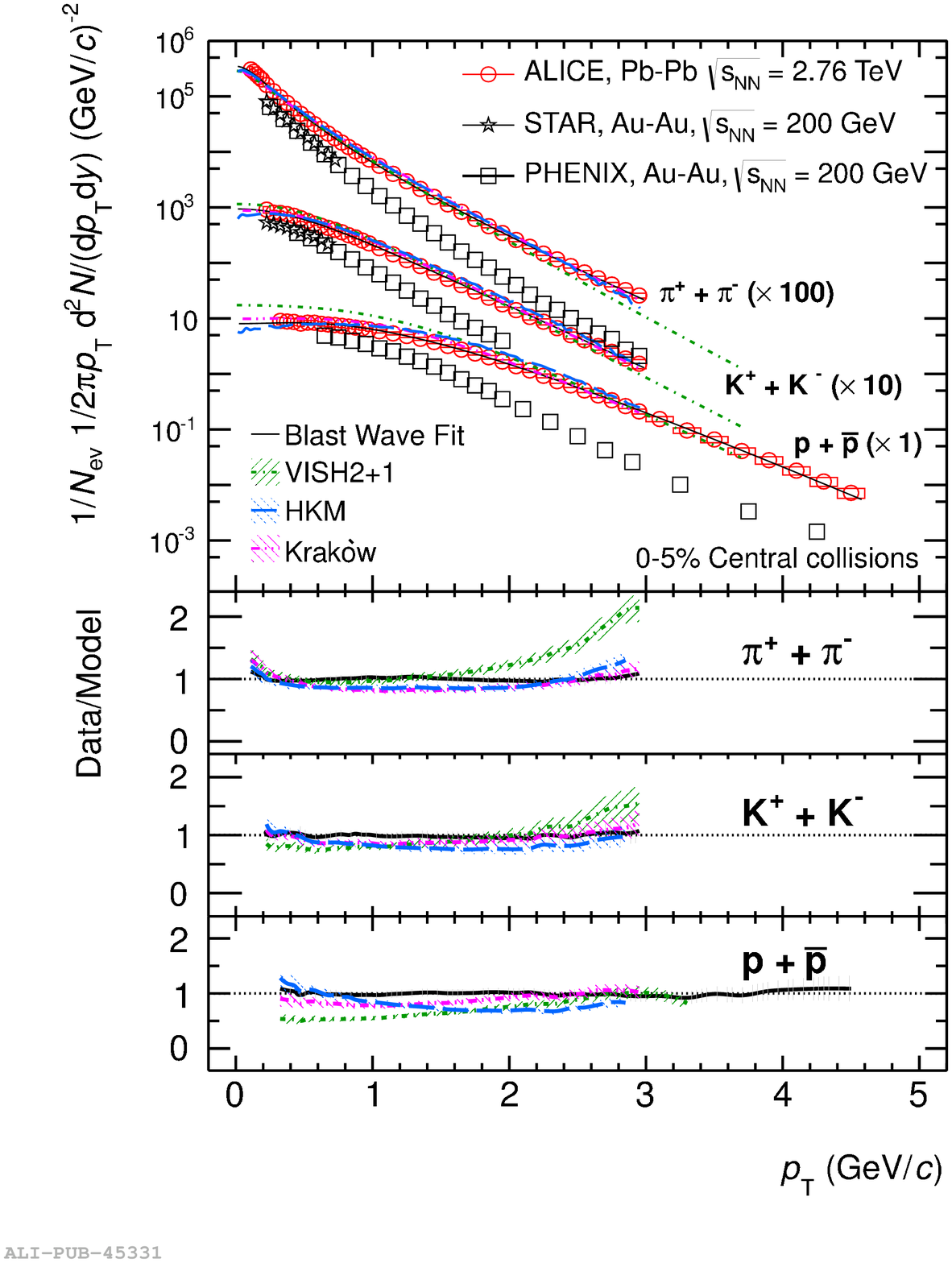}
\caption{Transverse momentum distributions of the sum of positive and negative pions, kaons and protons for central Pb--Pb collisions. The results are compared to RHIC data and hydrodynamic models.}
\label{fig:spectrapbpb}
\end{figure}

\begin{figure}[t]
\includegraphics[width=\linewidth,clip]{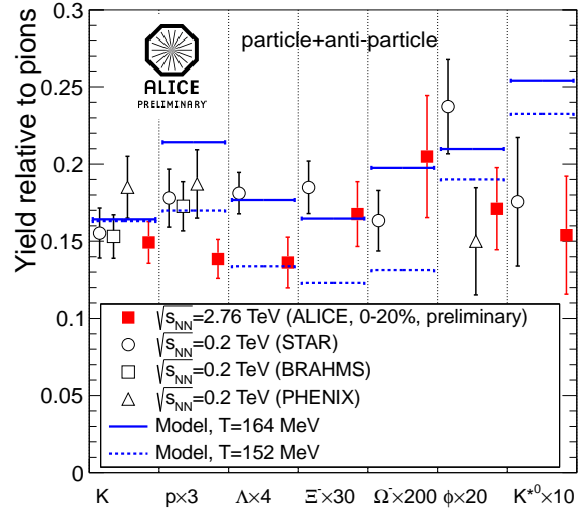}
\caption{Mid-rapidity particle ratios compared to RHIC results and predictions from thermal models for central Pb--Pb collisions at the LHC.}
\label{fig:ratiospbpb}
\end{figure}

\begin{figure}[t]
\includegraphics[width=\linewidth,clip]{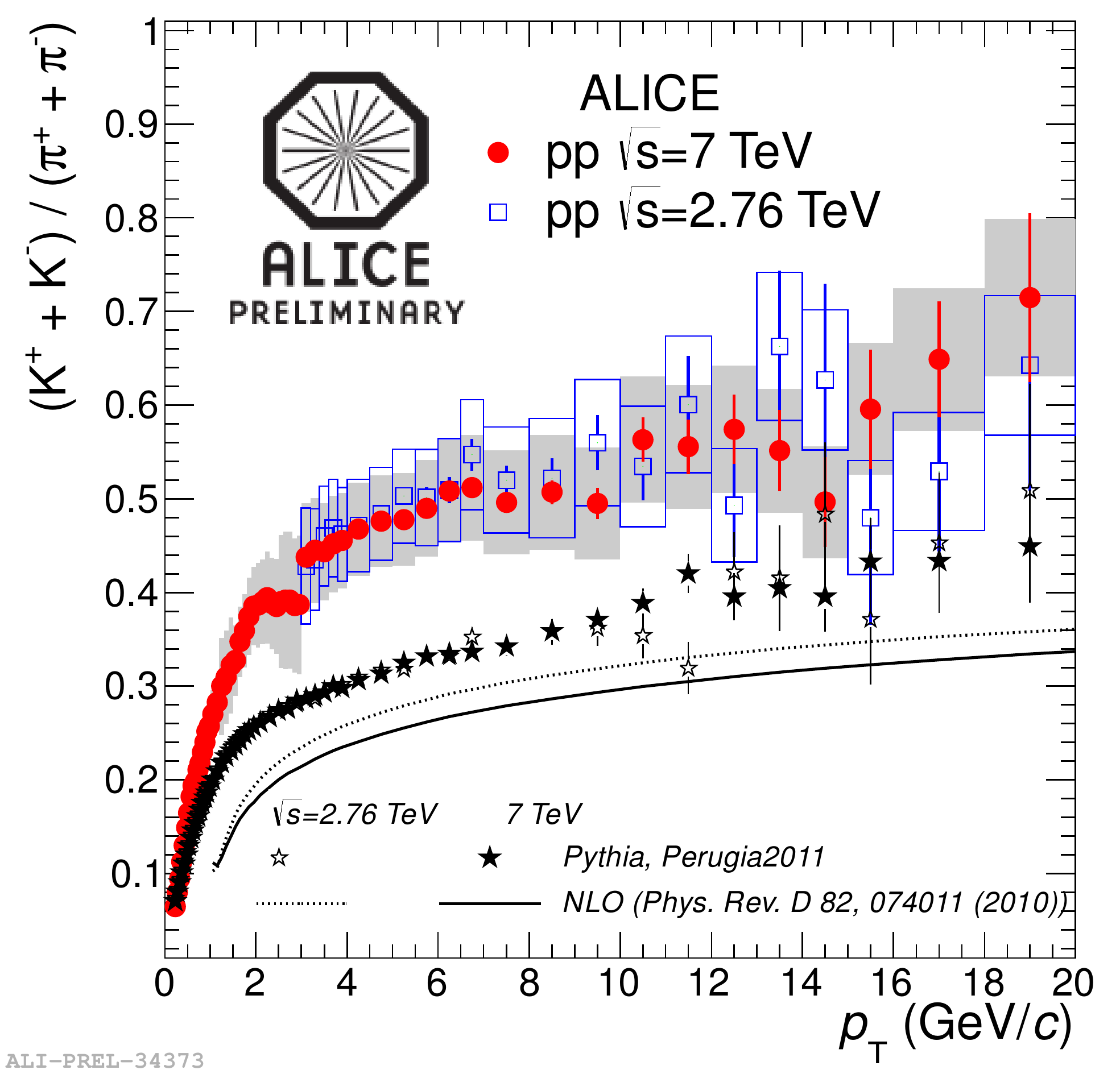}
\includegraphics[width=\linewidth,clip]{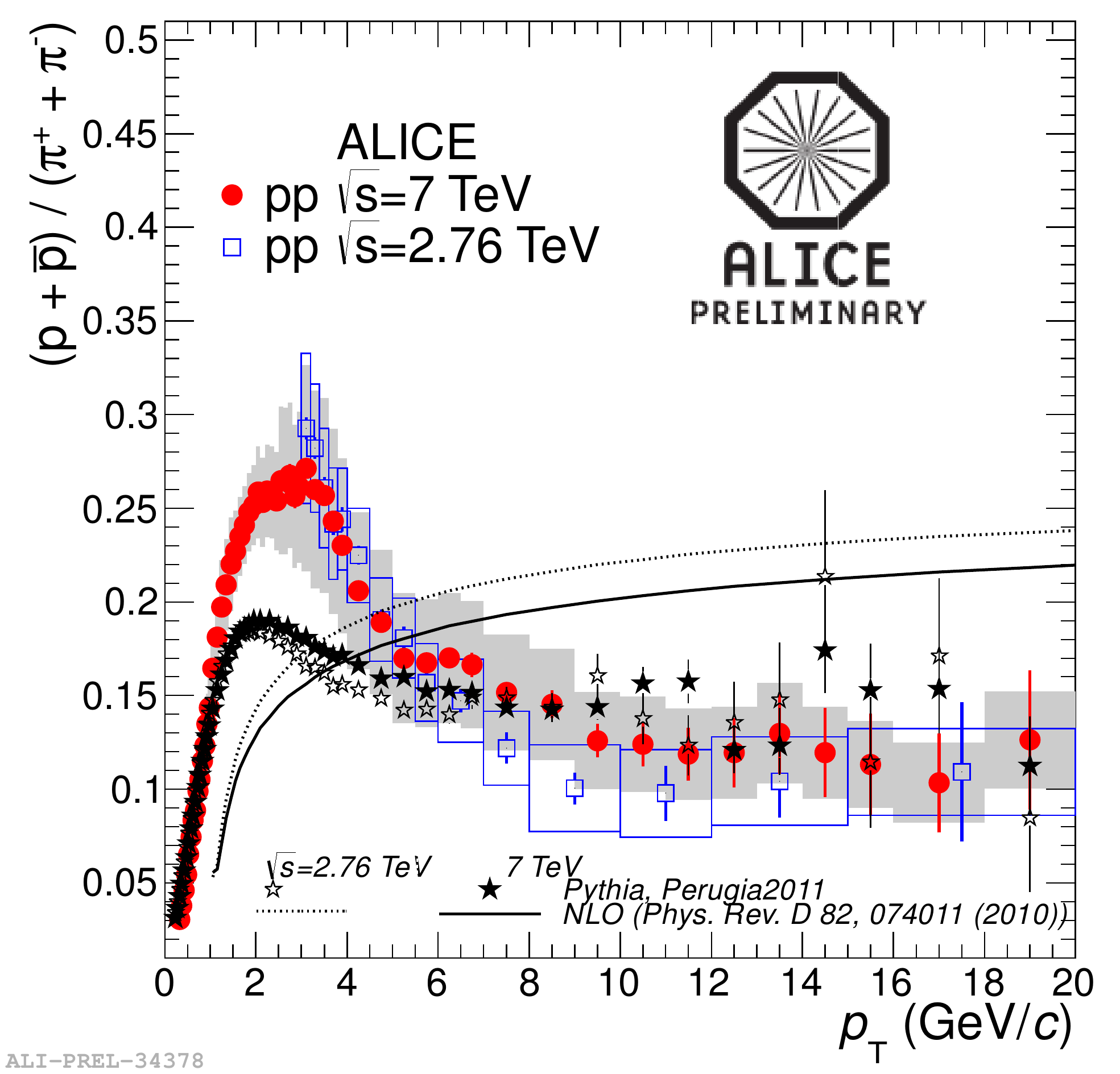}
\caption{$\rm K/\pi$ (top) and $\rm p/\pi$ production ratio in pp collisions compared with PYTHIA Monte Carlo predictions and NLO calculations.}
\label{fig:ratiospp}
\end{figure}

ALICE has measured the production yields of primary charged pions, kaon and (anti)protons in a wide momentum range. Primary particles are defined as prompt particles produced in the collision and all decay products, except products from weak decay of strange particles. The measurements have been performed both in proton-proton collisions at several centre-of-mass enegies ($\sqrt{s}$ = 0.9, 2.76 and 7 TeV) and in Pb--Pb collisions at $\sqrt{s_{\rm NN}}$ = 2.76 TeV as a function of collision centrality.

In Pb--Pb collisions the $p_{\rm T}$ distributions and yields are compared to previous results at RHIC and expectations from hydrodynamic and thermal models. The results obtained for central Pb--Pb collisions are shown in Figure~\ref{fig:spectrapbpb} and~\ref{fig:ratiospbpb} and are also reported in~\cite{:2012iu}. The spectral shapes are harder than those measured at RHIC, indicating a strong increase of the radial flow velocity with the centre-of-mass energy. The radial flow at the LHC is found to be about 10\% higher than at RHIC energy. While the K/$\pi$ integrated production ratio is measured to be in line with lower energy results and predictions from the thermal model, both the p/$\pi$ and the $\Lambda/\pi$ ratios are lower than at RHIC and significantly lower (a factor $\sim$ 1.5) than predictions. A possible explanation of these deviations from the thermal-model predictions may be re-interactions in the hadronic phase due to large cross sections for antibaryon-baryon annihilation~\cite{Steinheimer:2012rd,Becattini:2012sq,Pan:2012ne}.

The $p_{\rm T}$-dependent production of charged kaons and protons normalized to charged pions, respectively K/$\pi$ and p/$\pi$, are shown in Figure~\ref{fig:ratiospp} for pp collisions at $\sqrt{s}$ = 2.76 and 7 TeV where they are compared with theoretical model predictions. No energy dependence is observed in the data within the systematic uncertainties. The observed production ratios are not reproduced by NLO calculations. PYTHIA Monte Carlo generator underpredicts the proton-to-pion ratio at intermediate $p_{\rm T}$. Pion, kaon and (anti)proton production in Pb--Pb collisions were compared to that of proton-proton interactions and all show a suppression pattern which is similar to that of inclusive charged hadrons at high momenta ($p_{\rm T}$ above $\simeq$ 10 GeV/$c$)~\cite{OrtizVelasquez:2012te}. This suggests that the dense medium formed in Pb--Pb collisions does not affect the fragmentation. A similar conclusion can be drawn by observing the proton-to-pion ratio measured in Pb--Pb collisions: for intermediate momenta (3--7 GeV/$c$) it exibits a relatively strong enhancement, a factor 3 higher than proton-proton collisions at $p_{\rm T} \approx$ 3 GeV/$c$ and gets back to the proton-proton value at higher momenta ($p_{\rm T}$ above $\simeq$ 10 GeV/$c$)~\cite{OrtizVelasquez:2012te}. A similar observation is reported also for the $\rm \Lambda/K^{0}_{s}$ ratio and possible explanations include among those proposed so far particle production via quark recombination~\cite{Fries:2003vb}.

\subsection{Heavy-flavour production}

\begin{figure}[t]
\includegraphics[width=\linewidth,clip]{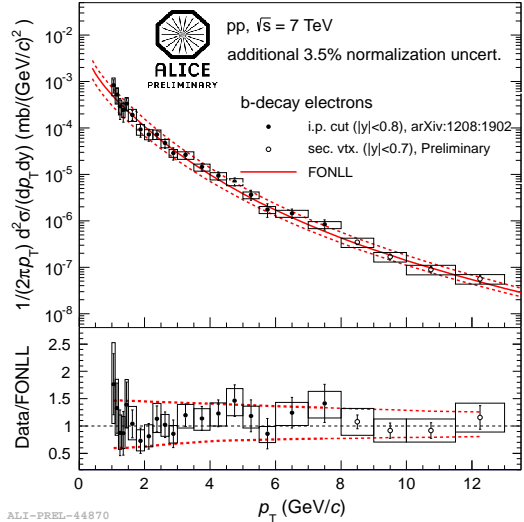}
\caption{$p_{\rm T}$-differential invariant cross sections of electrons from beauty hadron decays. The ratio of the data and the FONLL calculations are shown in in the bottom panel and the dashed lines indicate the calculation uncertainty.}
\label{fig:beauty}
\end{figure}

\begin{figure}[t]
\includegraphics[width=\linewidth,clip]{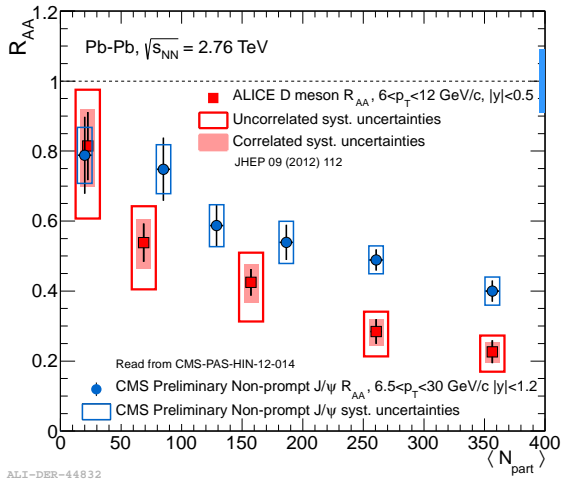}
\caption{Average D meson suppression versus number of participants compared with CMS non prompt $\rm J/\psi$ suppression.}
\label{fig:draa}
\end{figure}

ALICE has studied heavy-flavour production in pp and Pb--Pb collisions at the LHC. Their prodution in proton-proton collisions is a tool to test pQCD calculations in a new energy domain. The $p_{\rm T}$-differential production cross section of prompt charmed mesons (D$^{0}$, D$^{+}$, D$^{*+}$, D$^{+}_{s}$), heavy-flavour decay electron and muons in pp collisions at $\sqrt{s}$ = 7 and 2.76 TeV are reported in~\cite{Abelev:2012xe,Abelev:2012pi,Abelev:2012qh,ALICE:2011aa,:2012sx,:2012ik} and the measurements are well described by pQCD prediction. The production cross section of electrons from semileptonic decays of beauty hadrons was also measured at mid-rapidity ($\left| y \right| < 0.8$) in pp collisions at $\sqrt{s}$ = 7 TeV~\cite{:2012ic}. The results are shown in Figure~\ref{fig:beauty} together with the pQCD FONLL prediction.

Heavy-flavuor production studies in Pb--Pb collisions at $\sqrt{s_{\rm NN}}$ = 2.76 TeV were also performed~\cite{ALICE:2012ab,delValle:2012qw}. In particular, the relative production of heavy-flavour particles with respect to that in nucleon-nucleon interaction show a suppression of up to a factor of 5 at $p_{\rm T} \sim$ 10 GeV/$c$ in central Pb--Pb collisions (0-7.5\% centrality). The suppression is observed to be similar for the three studied decay channels, that is heavy-flavour electrons, heavy-flavour muons and prompt D mesons. The average D meson production suppression is shown in Figure~\ref{fig:draa} as a function of the number of participating nucleons and it is compared with the measurement of non-prompt $\rm J/\psi$ suppression performed by CMS~\cite{CMS:2012vxa}. The suppression of non-prompt $\rm J/\psi$ from B meson decays reflects the in-medium energy loss of $b$ quarks. The two measurements provide a first indication of different suppression for charm and beauty in central collisions, that is of different in-medium energy loss.

\subsection{Quarkonia}

\begin{figure}[t]
\includegraphics[width=\linewidth,clip]{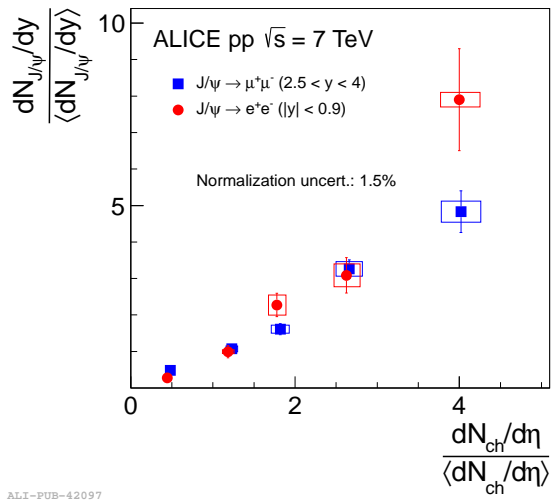}
\caption{$\rm J/\psi$ production yield ${\rm d}N_{\rm J/\psi}/{\rm d}y$ as a function of the charged particle multiplicity densities at mid-rapidity ${\rm d}N_{\rm ch}/{\rm d}\eta$. Both values are normalized by the corresponding pp minimum bias value.}
\label{fig:jpsimulti}
\end{figure}

\begin{figure}[t]
\includegraphics[width=\linewidth,clip]{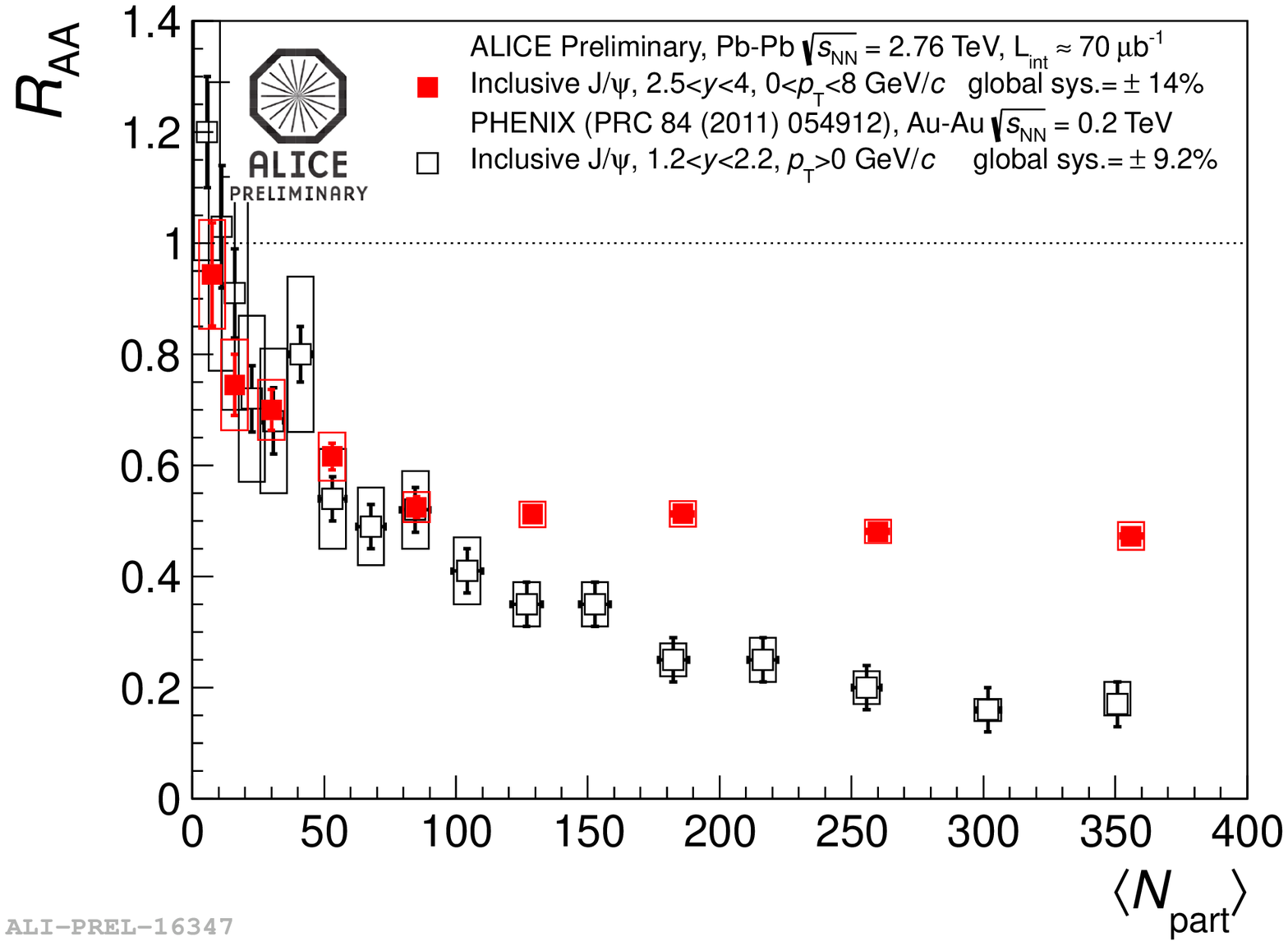}
\caption{Inclusive $\rm J/\psi$ suppression $R_{\rm AA}$ versus the number of participant nucleons $N_{\rm part}$ in Pb--Pb collisions, compared with results from PHENIX at RHIC.}
\label{fig:jpsiraa}
\end{figure}

ALICE has measured the inclusive $\rm J/\psi$ production in pp and Pb--Pb collisions at the LHC, down to zero $p_{\rm T}$~\cite{Aamodt:2011gj,Abelev:2012kr}. A unique study performed by ALICE in pp collisions is the determination of the inclusive $\rm J/\psi$ production yield as a function of the charged multiplicity at central rapidity. These results~\cite{Abelev:2012rz} show a linear increase of the yield with charged multiplcity which is not explained by theory and model predictions yet. A similar increase is observed both a forward and mid-rapidity, as shown in Figure~\ref{fig:jpsimulti}.

Results on inclusive $\rm J/\psi$ production in Pb--Pb collisions clearly indicate a saturation of the suppression at both central and forward rapidity moving towards central collisions, a phenomenon not observed at lower energy where the suppression increases as shown in Figure~\ref{fig:jpsiraa}. There is in fact a clear evidence for a smaller suppression at LHC with respect to RHIC energy~\cite{Abelev:2012rv}. Differential studies of the suppression versus collision centrality for various momentum bins seem to favour a scenario where (re)combination processes play a sizeble role~\cite{BraunMunzinger:2000px,Thews:2000rj}. The observed hint for a non-zero anisotropic production signal (elliptic flow) are in agreement with such a picture.

\section{Summary}

ALICE has obtained so far a wealth of physics results both from the analysis of proton-proton collision data and from the first two LHC heavy-ion runs. First results from a short pilot run with proton-lead have been already obtained and the coming dedicated p-Pb run at the beginning of 2013 will set the beginning of precision carachterization of the matter formed in heavy-ion collisions at the LHC. A clear detector upgrade strategy plan for the LHC luminosity upgrade has also been presented at this conference~\cite{ken}.

%
%
%

\end{document}